\documentclass{article}

\usepackage{graphicx}

\begin{document}

\title{Polariton Condensation and Lasing}
\author{David Snoke\\
Department of Physics and Astronomy\\
University of Pittsburgh, Pittsburgh, PA 15260, USA}
\date{}
\maketitle

\begin{abstract}
The similarities and differences between polariton condensation in microcavities and standard lasing in a semiconductor cavity structure are reviewed. The recent experiments on ``photon condensation'' are also reviewed.
\end{abstract}

Polariton condensation and lasing in a semiconductor vertical-cavity, surface-emitting laser (VCSEL) have many properties in common. Both emit coherent light normal to the plane of the cavity, both have an excitation density threshold above which there is optical gain, and both can have in-plane coherence and spontaneous polarization. Is there, then, any real difference between them? Should we drop the terminology of Bose-Einstein condensation (BEC) altogether and only talk of polariton lasing? 

The best way to think of this is to view polariton condensation and standard lasing as two points on a continuum, just as Bose-Einstein pair condensation and Bardeen-Cooper-Schreiffer (BCS)  superconductivity are two points on a continuum. In the BEC-BCS continuum, the parameter which is varied is the ratio of the pair correlation length to the average distance between the underlying fermions; BEC occurs when the pair correlation length, i.e., the size of a bound pair, is small compared to the distance between particles, while BCS superconductivity (which has the same mathematics as the excitonic insulator (EI) state, in the case of neutral electron-hole pairs) occurs when the pair correlation length is large compared to the distance between the particles. In the condensation-lasing continuum, the parameter which is varied is the ratio of the recombination time of the electron-hole pairs to the interaction time between pairs; in the exciton condensate limit, the recombination lifetime is essentially infinite compared to the time scale for interactions between the excitons, while in the case of lasing, the lifetime for recombination is short compared to the interaction time, and in fact, there may be no interaction between the pairs at all. The case of the polariton condensate lies somewhere between these two limits, since the polaritons are mixed states of excitons and photons. 

In these different limits, there are various properties of the BEC and standard lasing states that allow them to be distinguished. In this article we will review the connections between the two states and also discuss the experiments that show the differences.

\section{The state of matter in excitonic condensation and lasing}

The fact that exciton condensation and lasing are fundamentally similar can be seen by looking at the basic mathematics of exciton condensation.  In second quantization language, the creation operator for an exciton with center-of-mass momentum $K$ is (see, e.g., Ref.~\cite{snokebook}, Section 11.2.1),
\begin{eqnarray}
c^{\dagger}_K = \sum_k \phi(K/2-k) b^\dagger_{c,K-k}b^{ }_{v,-k},
\label{ccreate}
\end{eqnarray}
where $b^{\dagger}_{n,k}$ is the fermion creation operator for an electron in band $n$ with momentum $k$, and $\phi(k)$ is the orbital wave function for the electron-hole relative motion. In other words, the creating an exciton corresponds to creating a superposition of electron-hole pairs in the conduction band and valence bands. A Bose-Einstein condensate corresponds to a coherent state of excitons in a single $K$-state, which we will assume is $K=0$:
\begin{eqnarray}
|\alpha \rangle =  e^{-|\alpha|^2/2}\sum_n 
\frac{(\alpha c^{\dagger}_0)^n}{\sqrt{n!}}|0\rangle,
\label{coh}
\end{eqnarray}
where $\alpha = \sqrt{N}e^{i\theta}$ is the phase of the condensate, and 
$N$ is the number of excitons in the condensate. 

Let us look at this formalism in the limit of Frenkel excitons. Frenkel excitons correspond to excitons with electron and hole on the same lattice site, as opposed to Wannier excitons, in which the exciton and hole orbit each other at some distance. We write the fermion creation operator as the Fourier transform of a spatial creation operator for a single lattice site $i$:
\begin{eqnarray}
b^{\dagger}_{nk} = \frac{1}{\sqrt{N_s}} \sum_i e^{-ik\cdot r_i}b^{\dagger}_{ni},
\end{eqnarray}
where $N_s$ is the number of lattice sites. The creation operator (\ref{ccreate}) is then
\begin{eqnarray}
c^{\dagger}_K &=& \sum_k \phi(K/2-k) \frac{1}{N_s}\sum_{i,j} e^{-i(K-k)\cdot r_i}b^\dagger_{c,i}e^{-ik\cdot r_j}b^{ }_{v,j} \nonumber\\
&=& \frac{1}{N_s}\sum_{i,j} b^\dagger_{c,i}b^{ }_{v,j} e^{-iK\cdot r_i} \left(\sum_k \phi(K/2-k)  e^{ik\cdot (r_i-r_j)}\right).
\end{eqnarray}
The Frenkel limit corresponds to picking $\phi(k) = $ constant $ = 1/\sqrt{N_s}$, in which case the term in the parentheses, which is the real-space wave function for the electron-hole relative motion, equals $\sqrt{N_s}\delta_{ij}$. Putting this into our definition of $c^{\dagger}_K$ gives us
\begin{eqnarray}
c^{\dagger}_K &=& \frac{1}{\sqrt{N_s}}\sum_{i} e^{-iK\cdot r_i} b^\dagger_{c,i}b^{ }_{v,i} ,
\end{eqnarray}
i.e., a superposition of excitations at all lattice sites.

Creating a macroscopic number of $K=0$ excitons in one state involves terms of the form
\begin{eqnarray}
(c^{\dagger}_0)^n &=&   \left( \frac{1}{\sqrt{N_s}} \sum_{i}  b^\dagger_{c,i}b^{ }_{v,i}\right)^n 
\nonumber\\
&=& \frac{1}{N_s^{n/2}}\left(b^\dagger_{c,1}b^{ }_{v,1} + b^\dagger_{c,2}b^{ }_{v,2}+\ldots \right)^n.
\label{product}
\end{eqnarray}
This is a superposition of all possible ways to have $N$ excitons in $N_s$ lattice sites. 

The expectation value $ \langle \alpha | b^{\dagger}_{c,i}b^{ }_{c,i}|\alpha\rangle$ gives the probability of a lattice site $i$ being in the excited state. If we pick site $i$, then in the product (\ref{product}) there are $n$ terms which will create an exciton at that site. For each of these terms the number of ways of picking the remaining terms so as to not have double fermion creation operators (which vanish) is
$$
(N_s-1)(N_s-2)\ldots(N_s-n) = \frac{(N_s-1)!}{(N_s-n)!}.
$$
In the limit $N_s \gg n$, we ignore the possibility of double occupation and approximate
$$
\frac{(N_s-1)!}{(N_s-n)!}  \simeq N_s^{n-1},
$$
Each product term created in $|\alpha\rangle$ will have $n!$ matching terms in the term with the same order of $n$ in $\langle \alpha|$. The expectation value is therefore
\begin{eqnarray}
\langle \alpha | b^{\dagger}_{c,i}b^{ }_{c,i}|\alpha\rangle 
&=& e^{-N} \sum_n \frac{N^n}{(n!)^2} \frac{1}{N_s^n} \frac{n}{N_s} N_s^n n! \nonumber\\
&=& e^{-N}\sum_n \frac{N^n}{n!}  \frac{n}{N_s}\nonumber\\
&=& \frac{N}{N_s}.
\label{fermcoh}
\end{eqnarray}
In other words, the probability of a given lattice site being in the excited state is equal to the number of excitons in the condensate divided by the total number of lattice sites.If the number of excitons increases, at some point the possibility of double occupation becomes significant, and will cause this relation to become sublinear with $N$. This is known as ``phase space filling.''

The calculation for the dipole moment term $b^{\dagger}_{v,i}b^{ }_{c,i}$ is similar, except that this term changes the $n$ term in the coherent state sum to the $(n-1)$ term. We therefore have
\begin{eqnarray}
\langle \alpha | b^{\dagger}_{v,i}b^{ }_{c,i}|\alpha\rangle 
&=& e^{-N} \sum_n \frac{N^{n/2}}{n!}\frac{N^{(n-1)/2}}{(n-1)!}   \frac{e^{in\theta}e^{i(n-1)\theta}}{N_s^{n/2}N_s^{(n-1)/2}} \frac{n}{N_s}N_s^n (n-1)! \nonumber\\
&=& \sqrt{\frac{N_s}{N}}  e^{-i\theta} e^{-N} \sum_n \frac{N^{n}}{n!}   \frac{n}{N_s}    \nonumber\\
&=&  \sqrt{\frac{N}{N_s}}  e^{-i\theta} .
\label{fermcoh}
\end{eqnarray}
This is proportional to the amplitude of the condensate wave function. The total dipole moment of the solid is $N_s$ times this. This corresponds to a macroscopic radiating dipole moment which will emit coherent light.

Let us now switch to thinking about a standard laser. In the theory of lasing in an ensemble of independent two-level oscillators, we start with a single Bloch oscillator coupled to a coherent electromagnetic wave:
\begin{eqnarray}
( u_i + v_ib^\dagger_{c,i}b^{ }_{v,i})|0\rangle ,
\end{eqnarray}
where $|u_i|^2 + |v_i|^2 = 1$. The coefficient $v_i$ determines the degree of excitation, which is typically much less than unity. For an ensemble of identical oscillators, we repeatedly create identical states at all the different sites:
\begin{eqnarray}
| l \rangle = \prod_i (u+  vb^\dagger_{c,i}b^{ }_{v,i})|0\rangle.
\label{laserBCS}
\end{eqnarray}
The expectation value $\langle l | b^{\dagger}_{c,i}b^{ }_{c,i}|l\rangle$ is $|v|^2$, which we can set equal to $N/N_s$.

As shown in Ref.~\cite{snokebook}, Section 11.2.3 (and discussed earlier in Ref.~\cite{blatt}), the state (\ref{laserBCS}), which has the form of a BCS wave function, is equivalent to a coherent state of the form (\ref{coh}), where we equate
\begin{eqnarray}
\alpha \phi = \frac{v}{u}.
\end{eqnarray}
If we choose $\phi = 1/\sqrt{N_s}$ and $|v|^2 = N/N_s$, in the limit where $u \simeq 1$ when phase space filling is negligible, we have exactly the same state as our Frenkel exciton example, above, with probability of excitation of a single lattice site equal to $N/N_s$. 

The case of a coherent electromagnetic wave passing through a transparent medium with a dielectric constant (index of refraction) produces exactly the same state of matter. As with a laser, we model the system as an ensemble of two-level oscillators in phase with each other (see, e.g., Sections 7.1 and 7.3 of Ref.~\cite{snokebook}). In this case, the frequency of the wave is well below the electronic resonance, so that  the radiation from the polarized medium gives a phase shift to the electromagnetic wave, and a change in phase velocity, but no absorption or gain. 

We thus have three physical scenarios which all lead to a state of matter which consists of an ensemble of electronic excitations in phase with each other: excitonic condensation, lasing, and linear refraction of a wave in a dielectric medium. The first two involve spontaneous coherence, while the last has electronic coherence because it is driven by an external electromagnetic field. Some people have used the term ``driven condensate'' to refer to coherence of a medium driven by an external field, but this terminology is not very helpful, because it would have to also include every sound wave produced by a loudspeaker and every radio wave driven by an antenna. It is important to notice, though, that the polarization state of the underlying medium is fundamentally similar in all three cases; all are states of coherent polarization of matter. The differences arise in other properties, namely, how we get to the state, how stable the state is, the degree of inversion, and what is going on with other electronic states not participating in the coherence. 




\section{Condensation and classical waves}
\label{sect.GP}

We will discuss the differences of exciton-polariton BEC and standard lasing below, but before we do that, let us consider another connection. As discussed above, every BEC state is a coherent state of bosons of the form (\ref{coh}). A general property of coherent states with large $N$ is that they have exactly the same properties as classical wave states; in fact, in the modern understanding, all classical waves such as sound waves and electromagnetic waves are actually quantum-mechanical coherent states with high occupation number of bosons (phonons and photons, respectively). In the case of a standard BEC, the behavior of the coherent condensate is described by the Gross-Pitaevskii equation,
\begin{equation}
i\hbar\frac{\partial \psi}{\partial t} = 
-\frac{\hbar^2}{2m} \nabla^2\psi + U|\psi|^2\psi,
\label{GP}
\end{equation}
which becomes the Ginzburg-Landau equation if the time derivative is replaced by $E\psi$. This is also called the nonlinear Schr\"odinger equation, because it has exactly the same form as a single-particle Schr\"odinger equation with wave function $\psi$ and a nonlinear potential energy that is proportional to the square of the amplitude of the wave function, $|\psi|^2$. In the context of BEC, the wave function $\psi$ is normalized by $(1/L^d) \int d^d x\ |\psi|^2 = N$, where $N$ is the total number of particles in the condensate, and $d$ is the dimensionality. The Gross-Pitaevskii equation is often viewed as a heuristic assumption, but it can be justified fairly rigorously in terms of the underlying microsopics, as discussed in Ref.~\cite{snokebook}, Section 11.1.3. It is simply a consequence of the condensate being a coherent state, and the particles have a two-body interaction that leads to local potential energy proportional to the particle density. 

The fact that a condensate can be described by a classical wave equation with a nonlinear term has also led to other qualms that a polariton BEC is no different from a coherent classical optical system. We can see that the two have a fundamental connection by showing that we can also derive the Gross-Pitaevskii equation starting with Maxwell's equations. 

We start with Maxwell's wave equation in a nonlinear isotropic medium,
\begin{equation}
\nabla^2E = \frac{n^2}{c^2}\frac{\partial^2 E}{\partial t^2} + 4\mu_0\chi^{(3)}\frac{\partial^2 }{\partial t^2}|E|^2E,
\label{maxwell}
\end{equation}
where $\chi^{(3)}$ is the standard nonlinear optical constant, and we ignore frequency-mixing terms in the general $E^3$ nonlinear response.
We write a solution
\begin{equation}
E =\psi e^{-i\omega t},  
\end{equation}
where $\psi$ is an amplitude which may vary in time and space. We write this envelope amplitude suggestively as $\psi$ because we will see that it plays the same role as the matter-wave $\psi$ in the Gross-Pitaevskii equation. 

Keeping only leading terms in frequency (known as the slowly varying envelope approximation), we have for the time derivative of $E$,
\begin{eqnarray}
\frac{\partial^2 E}{\partial t^2} 
&\simeq&  \left(-\omega^2\psi -2i\omega\frac{\partial \psi}{\partial t}\right)e^{-i\omega t},  
\end{eqnarray}
and for the time derivative of the nonlinear term
\begin{eqnarray}
\frac{\partial^2 }{\partial t^2} |E|^2E 
&\simeq& 
-\omega^2|\psi|^2\psi e^{-i\omega t}.  
\nonumber
\end{eqnarray}

The standard polariton structure used a planar or nearly-planar cavity to give one confined direction of the optical mode. We therefore distinguish between the component of momentum $k_z$ in the direction of the cavity confinement and the momentum $k_{\|}$ in the two-dimensional plane perpendicular to this direction. We therefore write
\begin{equation}
\psi = \psi(\vec{x})e^{i(k_{\|}\cdot\vec{x}+k_zz)}.
\end{equation}
The full Maxwell wave equation (\ref{maxwell}) then becomes
\begin{eqnarray}
&& (-(k_z^2+k_{\|}^2)\psi + \nabla_{\|}^2\psi)\nonumber\\
&&\hspace{1cm}
=(n/c)^2\left(-\omega^2\psi -2i\omega\frac{\partial \psi}{\partial t}\right)-4\mu_0\chi^{(3)}
\omega^2|\psi|^2\psi  
.\nonumber
\end{eqnarray}
Since $\omega^2 = (c/n)^2(k_z^2 + k_{\|}^2)$, this becomes
\begin{eqnarray}
&& \nabla_{\|}^2\psi 
=(n/c)^2\left(-2i\omega\frac{\partial \psi}{\partial t}\right)-4\mu_0\chi^{(3)}
\omega^2|\psi|^2\psi  .
\label{opticalGP}
\end{eqnarray}

Near $k_{\|} = 0$, we can approximate
\begin{equation}
\hbar\omega  =\hbar(c/n) \sqrt{k_z^2 + k_{\|}^2} \ \simeq  \ \hbar(c/n)k_z \left(1+\frac{k_{\|}^2}{2k_z^2}\right)
\equiv \hbar\omega_0 + \frac{\hbar^2k_{\|}^2}{2m},
\end{equation}
which gives an effective mass for the photon motion in the plane. This does not take into account the renormalization of the effective mass due to the coupling with the exciton states, which typically makes the effective mass about a factor of two heavier. To account for this, we would have to take into account the dependence of the index of refraction $n$ on frequency due to the exciton resonance, as discussed in Ref.~\cite{snokebook}, Section 7.4.
Neglecting this correction, for the first term on the right-hand side, we approximate
\begin{equation}
\omega \simeq \omega_0 = \frac{m(c/n)^2}{\hbar},
\end{equation}
so that we have
\begin{equation}
i\hbar\frac{\partial \psi}{\partial t} = 
-\frac{\hbar^2}{2m} \nabla_{\|}^2\psi - \frac{2\mu_0\chi^{(3)}(\hbar\omega)^2}{m}|\psi|^2\psi,
\label{GP1}
\end{equation}
which we can rewrite as
\begin{equation}
i\hbar\frac{\partial \psi}{\partial t} = 
-\frac{\hbar^2}{2m} \nabla_{\|}^2\psi + U|\psi|^2\psi.
\label{GP2}
\end{equation}
This is the Gross-Pitaevskii equation, or nonlinear Schr\"odinger equation. Note that although the Maxwell wave equation is second order in the time derivative, this equation is first order in time derivative, as in a typical Schr\"odinger equation.

One useful result of this derivation is that we can write the nonlinear $\chi^{(3)}$ value in terms of  experimentally measured parameters. To do this we first need to calibrate the electric field amplitude in terms of the density of particles.  From basic photon theory (see Ref.~\cite{snokebook}, Section 4.4) the number of photons is related to the electric field amplitude by
\begin{equation}
N = \frac{\epsilon_0 V}{2\hbar\omega}E^2,
\end{equation}
which for our two-dimensional system can be written as
\begin{equation}
\psi^2 =    \frac{2\hbar\omega}{\epsilon_0 d} \frac{N}{A},
\end{equation}
where $A$ is the area and $d$ is the effective thickness.
If we switch $\psi$ in (\ref{GP2}) from standing for electric field amplitude to standing for a number density wave function, the nonlinear term therefore corresponds to a potential energy linearly proportional to the density, with constant of proportionality equal to 
$$
U\  = \frac{2\mu_0|\chi^{(3)}|(\hbar\omega)^2}{m} \frac{2\hbar\omega}{\epsilon_0 d}  .
$$
We can use this result to equate the measured coefficient of linear shift of the energy of the polaritons, measured from the shift of the spectral line, with an effective $\chi^{(3)}$ value. Typical values for polariton condensates are 1 meV of shift for a density of $10^9$ cm$^{-2}$.  Using a typical microcavity length of 400 nm and material constants for GaAs, we obtain
\begin{eqnarray}
\frac{|\chi^{(3)}|} {\epsilon_0} &=& \frac{Umd}{4\mu_0(\hbar\omega)^3} \\
&\simeq & 3\times 10^{-14}\ {\rm m}^2/{\rm V}^2,
\end{eqnarray}
which is equivalent to $2\times 10^{-6}$ esu.
By comparison, typical nonlinear coefficients for strongly nonlinear media are around three orders of magnitude lower \cite{nonlin}. This is one of the main appeals of polariton condensates as optical systems---they have world-record nonlinearities, allowing, for example, optical parametric oscillator (OPO) behavior with continuous laser pump powers of a milliwatt or less, compared to standard broadband OPO systems which require pulsed lasers with instantaneous power at least six orders of magnitude higher. The tradeoff is that polariton systems only work for a narrow range of wavelengths; the huge nonlinearity is obtained by operating near a sharp exciton resonance. This can be done with resonances of single atoms, also, but to do that requires ultra-high vacuum and extremely good laser frequency stability. 

The fact that that polariton condensate Gross-Pitaevskii equation maps to the nonlinear Maxwell wave equation in this system should not be cause to doubt the validity of the BEC description. Every condensate is described by a classical wave equation, because the Gross-Pitaevskii equation is a classical wave equation. It does tell us, however, that some experimental results with polariton condensates do not intrinsically distinguish between condensation and lasing, or more generally, a coherent optical field. For example, quantized vortices are fundamentally a result of having a single-valued, coherent wave. Since this occurs also in the optical modes of a laser, quantized vortices can also be seen in a standard laser or VCSEL \cite{laservort}. In the same way, in-plane phase coherence of the wave function is also seen in spatially extended laser modes from a VCSEL \cite{botao}. However, the central core radius of a vortex of a polariton BEC, in which the condensate is depleted, is determined by the healing length of the condensate $\xi$, which is determined by the interaction strength of the particles---stronger interaction gives shorter healing length (See Ref.~\cite{snokebook}, Section 11.1.4). In vortices of laser modes, the vortex structure is determined by the spatial dimensions of the laser modes.


\section{Differences between polariton condensation and lasing}
\label{sect.diff}

\begin{figure} 
\begin{center}
\includegraphics[width=0.95\textwidth]{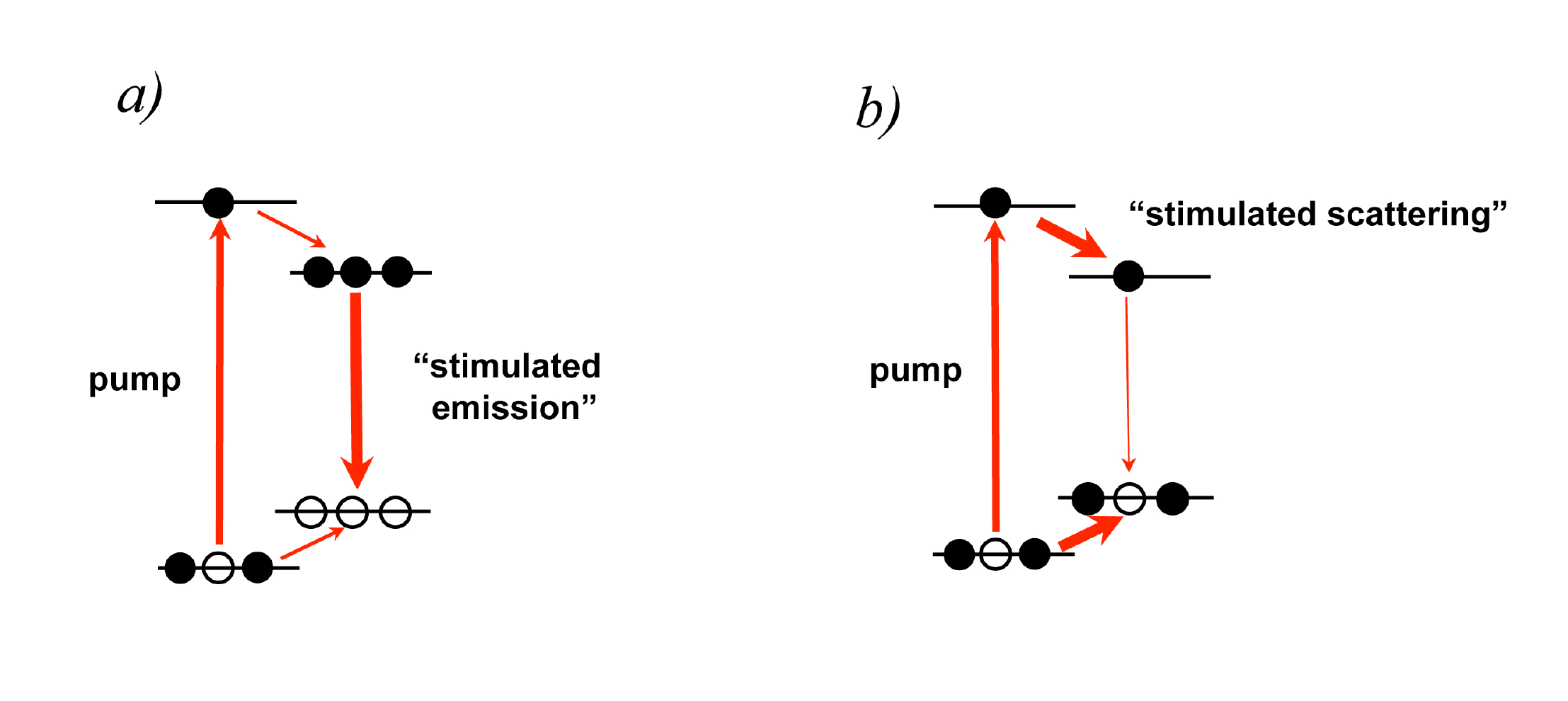}
\caption{a) Pumping and emission scheme for a standard semiconductor laser. b) Pumping and emission for an excitonic condensate. The emission rate can be low, with no optical gain in the emission process. Gain occurs in the redistribution of the excited carriers.}
\label{laser1}
\end{center}
\end{figure}
To see the difference between a polariton condensate coherent state and a standard laser, we start by examining how these states arise. Figure~\ref{laser1}(a) shows the standard band picture of a semiconductor laser. Electrons from the valence band are pumped high up into the conduction band through what is typically an incoherent process, and then both the electrons and holes created by the pumping fall in energy to near the band edges (holes float upwards as they lose energy, in this picture). This energy relaxation is an incoherent process, so that even if a coherent pump was used to excite the carriers, by the time they arrive in  the lowest energy states, they have lost all memory of the pump laser coherence. Fundamentally, the semiconductor laser is no different from an ensemble of isolated atoms, e.g. atoms in a gas or dopant atoms in a transparent matrix, with two occupied levels and two unoccupied levels.  The pumping of the electrons into the upper states leads to {\em inversion}, that is, higher probability of carriers in the upper state than in the lower state of the two levels involved in the light emitting transition (so-called ``negative temperature'').  This leads to a spectral range for electromagnetic waves with {\em gain}, that is, amplification of coherent electromagnetic waves. As shown in Ref.~\cite{snokebook}, Section 11.3.1, a simple model of a laser as a homogeneous ensemble of pumped two-level atoms leads to the result
\begin{eqnarray}
\frac{\partial \psi}{\partial t} = \frac{\omega}{2\epsilon} (A\psi - B|\psi|^2\psi),
\label{lasegain}
\end{eqnarray}
where $\epsilon$ is the dielectric constant and $A$ and $B$ are two terms which depend on the pumping rate and oscillator strength; $A$ is positive for net pumping rate exceeding the spontaneous emission rate, and $B$ has the same sign as $A$. If the system is initially incoherent, but there is some small fluctuation that gives a coherent seed, this coherence will be amplified by the $A$ term until there is a macroscopic coherence. The amplification will eventually be limited by the nonlinear $B$ term, which arises from the physical constraint that the upper level of the two-level atoms cannot have higher than unity occupation, and Rabi flopping will depopulate the upper level if the coherent field grows too strong. This is equivalent to saying that there is a nonlinear term due to {\em phase space filling}, that is, filling of the upper states due to the Pauli exclusion principle for the fermion electrons. 

One can see from this that the onset of lasing is a type of spontaneous symmetry breaking, since we cannot predict the exact value of the complex amplitude of the seed that will be amplified by (\ref{lasegain}). The same type of spontaneous symmetry breaking occurs in Bose-Einstein condensation.
A calculation to be presented elsewhere \cite{snokebeconset} gives for the complex amplitude $ \langle a_k \rangle$ of a condensate
 \begin{eqnarray}
\frac{d}{dt} \langle a_k \rangle &=& \langle a_k \rangle \frac{2\pi }{\hbar} 2U^2 \sum_{k_2,k_3}  \left[
\langle \hat{N}_{k_3}\rangle \langle \hat{N}_{k_4}\rangle (1+ \langle \hat{N}_{k_2}\rangle) \right. 
\nonumber\\
&&\left. -\langle \hat{N}_{k_2}\rangle(1+ \langle \hat{N}_{k_3}\rangle)(1+\langle \hat{N}_{k_4}\rangle) \right]   \delta(E_{k}+E_{k_2}-E_{k_3}-E_{k_4}),\nonumber\\
\label{bosegain}
 \end{eqnarray}
where the $\langle \hat{N}_k\rangle$ are the average occupation numbers of the excited $k$-states, the same occupation numbers that go into the quantum Boltzmann equation for the increase of the population in the condensate (see Ref.~\cite{snokebook}, Section~4.8). The first term is normally small relative to the second, since it involves a product of two occupation numbers while the second involves only one, but when the number of particles in low-energy states becomes large, it can lead to gain of the coherent amplitude of the condensate. (The topic of the onset of phase coherence in BEC has been a longstanding problem in the theory of BEC, addressed by numerous authors (e.g., Refs.~\cite{stoof,scully,zoller}.) It is not essential that a collisional interaction be the means of establishing condensation; condensation can also occur by emission and reabsorption of phonons (see Ref.~\cite{moskbook}, Section 8.2.3). But in every case, the stimulated scattering $(1+\langle \hat{N}_k\rangle)$ factors, which connect the occupation of one particle state to another, drives the onset of the phase coherence. As illustrated in Fig.~\ref{laser1}(b), the gain in the system occurs in the coupling of the continuum of excited states to the emitting states, not in the emission process itself. 

These two different relations, (\ref{lasegain}) and (\ref{bosegain}), tell us a lot about both the similarities and the differences between lasing and excitonic condensation. We see that both are examples of spontaneous coherence. In the spontaneous symmetry breaking of a laser, the control parameter is not the temperature $T$, as the transition is not a thermodynamic phase transition; instead the control parameter is the incoherent pumping rate which goes into the gain constant $A$. Nevertheless, spontaneous symmetry breaking is not unique to BEC. 

The {\em onset} and {\em stability} of standard lasing arise from the net gain from incoherent pumping, and the coherence arises from the stimulated emission of photons, which is taken into account in (\ref{lasegain}) by the $A\psi$ term. This implies that a standard laser must have {\em inversion}, and the stimulated emission occurs {\em only in the coherent state}-- whatever state has many photons will be amplified if there is gain. In a standard laser, that photon state is defined by the mirors of the cavity, which recirculates photons to suppress the loss rate. Electronic excitations not resonant with that state will undergo transitions to pump that state only incoherently. 

\begin{figure}
\begin{center}
\includegraphics[width=0.8\textwidth]{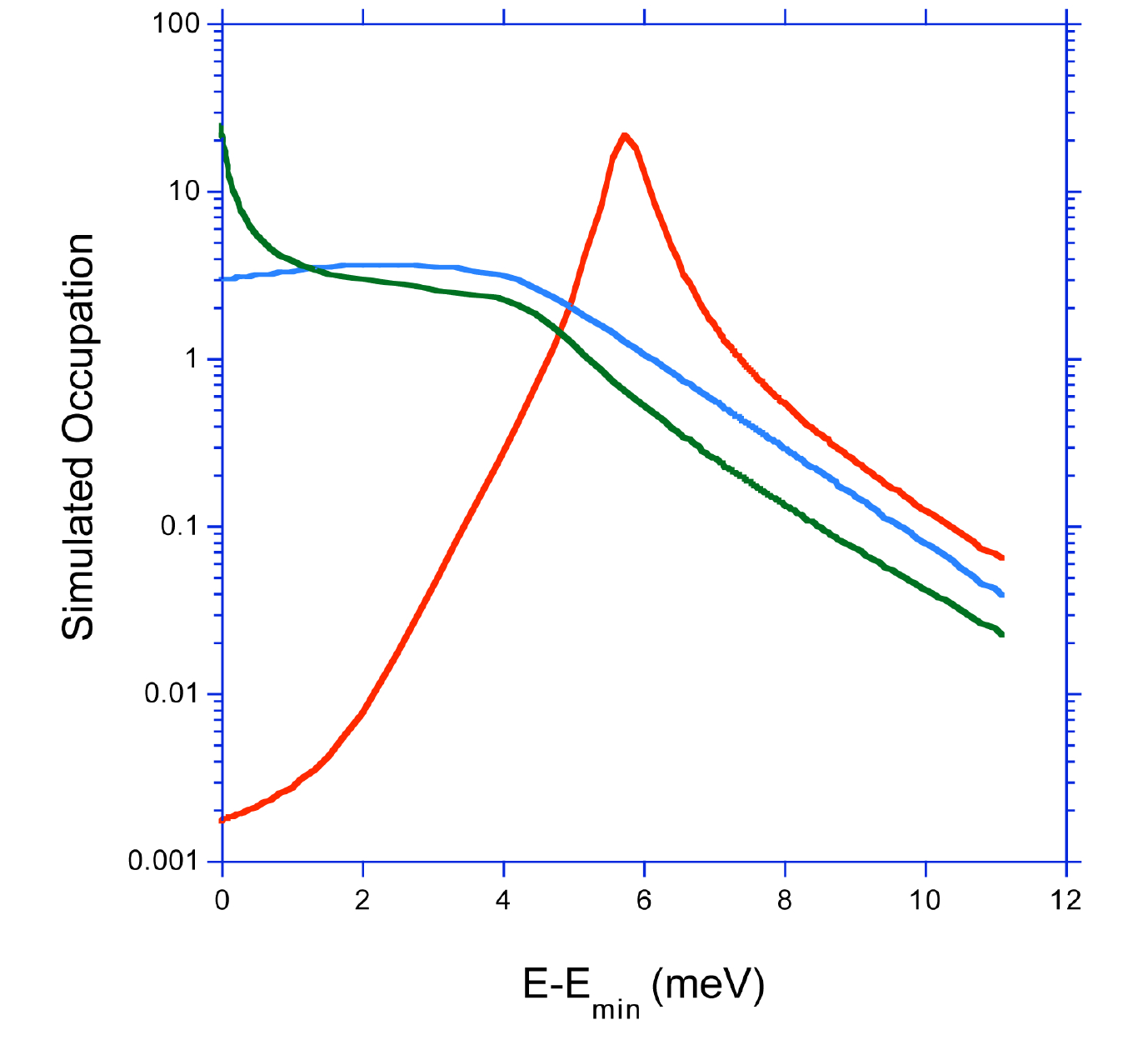}
\end{center}
\caption{Energy distribution of polaritons in a GaAs microcavity structure from numerical solutions of the quantum Boltzmann equation in steady state for the parameters of the experimental data. Red, lower curve: when only interactions with phonons are accounted for. Above 6 meV the states are essentially exciton states, with a Maxwell-Boltzmann distribution. The ``polariton bottleneck'' at 6 meV occurs because the phonons cannot efficiently couple exciton states to polariton states. Blue curve: polariton-polariton interactions included, but final-states $(1+N_k)$ factors not included. Green curve: full calculation including final states $(1+N_k)$ factors.  There is not only a peak ak $E=0$, there is also a pileup of particles in low-momentum states.  From Ref.~\protect\cite{hartwell}.}
\label{pileup}
\end{figure}

In a polariton condensate, as with any condensate, the onset and stability of the coherent amplitude come fundamentally from the {\em interaction of the particles in nearby states}. The two terms which amplify the coherence in (\ref{bosegain}) come from the occupation numbers of the non-condensed states. If all of the occupation numbers $\langle \hat{N}_{k_2}\rangle$, $\langle \hat{N}_{k_3}\rangle$, and $\langle \hat{N}_{k_4}\rangle$ are all small compared to unity, then there will always be net dephasing. If $\langle \hat{N}_{k_3}\rangle$ and $\langle \hat{N}_{k_4}\rangle$ are large compared to unity, however, while $\langle \hat{N}_{k_2}\rangle$ is small, then there will be amplification. Onset of condensation therefore requires not just that the condensate state itself is highly occupied, but that nearby states also have large (though not necessarily macroscopic) occupation. The coherence of the system is shared as a collective property of particles in many states.  Figure~\ref{pileup} shows how the states near to the condensate state in a polariton system are also highly occupied. 

In a condensate of particles with finite lifetime, there is a pump that incoherently creates new particles to compensate for the loss of the particles, but the stimulated process that gives rise to coherence is in the scattering of these incoherently created particles with each other, not in gain in the recombination process. Inversion is not only not necessary for polariton condensation, it actually kills polariton condensation, because inversion corresponds to Pauli state filling of the bands, which prevents pairing of the carriers into bosonic excitons, and thus shuts off the process of bosonic stimulated scattering which leads to condensation. 

One can think of the two limits of the BEC-lasing continuum in terms of what drives the coherence. In the case of lasing, we can imagine starting with a very small coherent electromagnetic field which drives the two-level oscillators, similar to the way that an electromagnetic wave drives the oscillators which give the dielectric response of the medium. This small coherent electromagnetic wave drives a small coherent polarization of the medium, which then radiates, and this radiation is amplified by the gain due to the inversion. 

In an excitonic BEC, the coherent polarization of the medium drives the electromagnetic radiation. The coherence in the polarization of the medium leads to emission of coherent radiation, but would still occur even if there were no radiation; which can be the case, for example, when the emission of radiation is forbidden by the symmetry of the semiconductor bands \cite{cu2o} or by a barrier between the electrons and holes, as in the case of excitons in double quantum wells \cite{DQW}. We imagine starting with a small coherent polarization of the medium, which then is amplified by pulling in carriers from the incoherent excited population of the medium, and then this amplified coherent polarization can emit radiation.

In other words, in both systems there are electromagnetic degrees of freedom and electronic degrees of freedom, but in a laser the coherence of the electronic states of the medium is driven by the electromagnetic coherence, as in a linear coherent wave passing through a dielectric medium. In an excitonic condensate, the coherence of the medium arises first, and coherent radiation is emitted by this polarization as a byproduct in some cases. The case of an exciton-polariton condensate is a middle ground, in which electromagnetic and electronic degrees of freedom participate equally. As we will see below, the electronic degrees of freedom still play an important role.

\section{Experimental differences between condensation and lasing}
\label{sect.expt}

A standard exciton-polariton microcavity structure used for condensation is different from a laser in having {\em strong coupling} between exciton and photon states. Strong coupling is defined as a Rabi splitting between the upper and lower polariton states which is large compared to the line with of either line. (For a review of the basic properties of microcavity polaritons, see Ref.~\cite{kavokinbook}.) The term ``Rabi frequency'' for the splitting between the states is no accident---it is equal to the standard Rabi frequency $\omega_R = (e\langle p \rangle/m\hbar\omega_0)E_0$ in a two-level oscillator, where $\langle p \rangle$ is the dipole matrix element coupling the two states, when $E_0$ is the electric field amplitude of a single photon of frequency $\omega_0$ in the cavity, and $\omega_0$ is the exciton resonance. 

When the system is in strong coupling,  the polariton is the proper eigenstate, with an equal, or near-equal, superposition of an exciton and a photon. Thus, if spontaneous coherence occurs in this strong coupling regime, we may say that the electronic polarization is playing an important role in driving the transition, as in an excitonic condensate. 

In many experiments, the system may be in strong coupling at low excitation density but revert to weak coupling at high density. The reason is that the dipole matrix element $\langle p \rangle$ is in general a function of carrier density, and can become greatly reduced at high density due to Pauli phase space filling and carrier screening---essentially, the excitons no longer are well-defined bound states at high density, so that the system becomes a gas of uncorrelated electrons and holes, i.e. a plasma, and the dipole coupling between photons and free electrons and holes is much less than the coupling of photons to excitons. The transition from excitons to free carriers is sometimes called the ``Mott transition,'' although the nature of this exciton-plasma transition in semiconductors is of quite different nature from that considered in cold atom gases and in doped solids \cite{mott,manzke}. 

If the system reverts to weak coupling, this will result in closing of the gap between the upper and lower polaritons at the zero detuning point (the point at which the excitons and photons have exactly the same energy). It is not always easy to know when this has happened, however. If the bare exciton and photon states are not perfectly equal in energy, then there will be a splitting between them just due to the energy difference of the bare states. This energy gap will not close down when the system goes into weak coupling. A better test of strong coupling is to measure the effective mass of the lower polaritons via angle-resolved far-field emission (this has been done in several experiments). In strong coupling, the effective mass of the polaritons will be twice the mass of the bare cavity photon mode \cite{berman}. 

One way to allow higher polariton density without reverting to weak coupling, which has been adopted by many groups, is to use multiple quantum wells in a cavity instead of just one. This greatly reduces the amount of phase space filling, since the exciton component of any one polariton is shared among many wells. 
Typical samples now use 12 or 16 wells placed at antinodes of the confined cavity photon mode \cite{balili,bloch,yamamoto}.

\begin{figure}
\begin{center}
\includegraphics[width=0.85\textwidth]{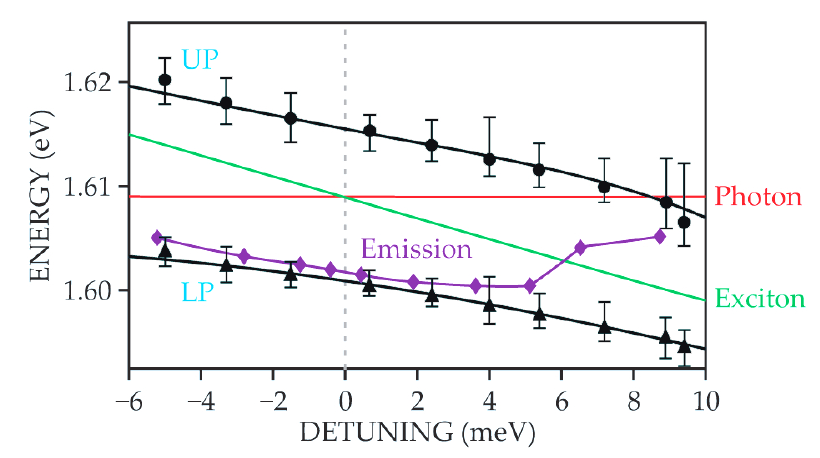}
\end{center}
\caption{Black circles: energy of the upper polariton (UP) and lower polariton (LP) from reflectivity data in a GaAs microcavity, as a function of detuning as the exciton state is shifted by applied stress. Purple diamonds: the energy of the coherent emission from the sample as a function of stress. At detuning less than +4 meV, the coherent emission follows the exciton shift, showing the system is still in the strong coupling limit. The green and red lines are the bare exciton and photon modes, determined by the fit to the reflectivity data shown as the black lines.  The line connecting the purple diamonds is a guide to the eye. From Ref.~\protect\cite{balilishift}.}
\label{fig.stress}
\end{figure}
In such samples it is possible to observe a clear distinction between polariton condensation and lasing. 
 Figure~\ref{fig.stress} shows the emission energy from polaritons in a microcavity in which the exciton energy was shifted using stress to tune the exciton resonance frequency, while leaving the photon cavity frequency essentially unchanged. As seen in the reflectivity data, the lower polariton mode follows the exciton shift with stress, since it has an excitonic component when there is strong coupling.  At high density, when the polariton condensate appears, the coherent emission still shifts with stress to follow the exciton mode. If the system were in a lasing mode with weak coupling, the emission would occur at the cavity photon mode, indicated by the dashed line in Fig.~\ref{fig.stress}. (Actually, at high density this cavity mode is slightly red shifted from its low-density value, due to a density-dependent change of the index of refraction \cite{blochred}, but this red shift is unaffected by the stress if the system is in weak coupling.) As seen in this figure, at very large detuning, the coherent emission finally jumps up to near the bare cavity photon frequency, as the system reverts to weak coupling, and the system is at the point in standard VCSEL lasing mode. 

The excitation density required to get coherent emission is not the same in the polariton condensate state and the lasing state. The polariton condensate in general requires much lower excitation power, because it does not require inversion, as discussed above. In the sample used for the data of Fig.~\ref{fig.stress}, the 
threshold for polariton condensation in the limit is about 6 times smaller than the threshold for lasing at large positive detuning at the same place on the same (recall that via stress tuning, the detuning can be varied for the same spot of the sample.)  Ref.~\cite{yamamoto} showed that polariton condensation could have a threshold as much as a factor of 20 lower than the lasing threshold in the same sample, when the detuning was changed by moving the laser spot to a different place in the sample with different cavity thickness. 

\begin{figure}
\begin{center}
\includegraphics[width=0.95\textwidth]{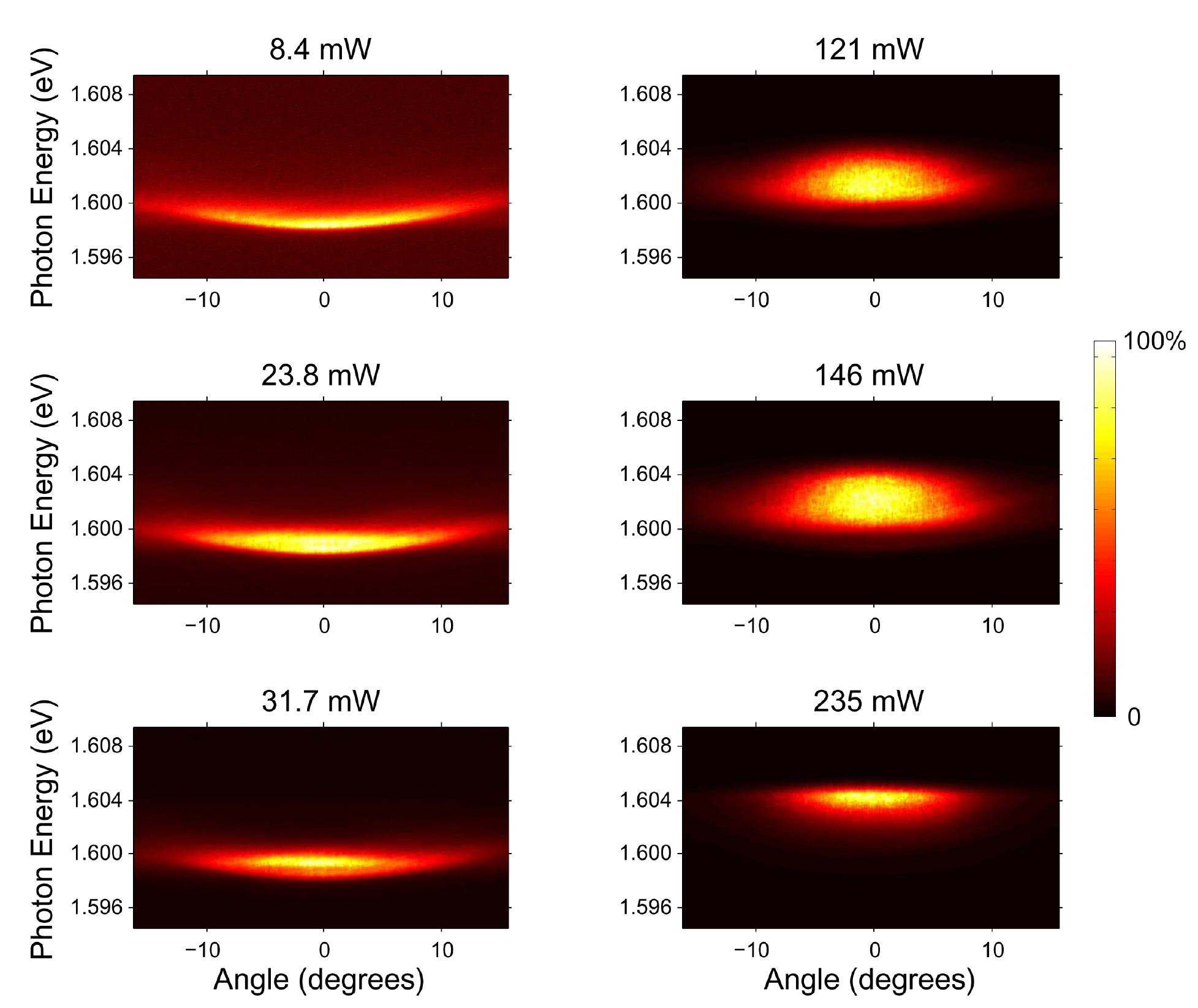}
\end{center}
\caption{Angle-resolved photon emission from a GaAs microcavity sample (which gives the momentum and energy distribution, i.e. spectral function times the occupation number), for several excitation pump intensities. Polariton condensation is at 31.7 mW, with energy and momentum narrowing; standard lasing is seen at 235 mW. From Ref.~\protect\cite{JAP}.}
\label{fig.series}
\end{figure}

Since the lasing threshold is higher in power, it is possible to push the system from polariton condensation to standard lasing simply by turning up the excitation power. Figure~\ref{fig.series} shows a sequence of images which give the angle-resolved spectrum of the polariton emission from the same microcavity sample used for Fig.~\ref{fig.stress} when the stress is chosen to give nearly zero detuning, and the excitation power is increased.  The angle of emission of the externally emitted photons maps directly to the momentum of the polaritons in the plane of their motion. 

Comparing the 23.8 mW data to the 8.4 mW data, one sees that at first, as the power is increased, the polariton spectrum shifts to the blue, and broadens. Both of these effects are expected for particles with repulsive interactions. Because these spectra are time averaged, there is also a contribution to the broadening if the blue shift fluctuates in time due to laser power fluctuations.  The 31.7 mW data shows the spectrum near the polariton condensation threshold. The spectrum has narrowed in energy, and also has become narrower in momentum space. The energy width here is limited by instrumental resolution; Ref.~\cite{skolT2} has recently showed that extreme narrowing, corresponding to coherence times of more than 200 ps, can be measured in this type of system when a vert stable laser with low intensity fluctuations is used. 

If the power is increased further, the spectrum broadens again, and shifts even more strongly to the blue. 
Both of these effects are related to phase space filling at high density. As discussed above, phase space filling reduces the coupling, and therefore reduces the Rabi splitting between the upper and lower polaritons, which shifts the lower polariton up. The increased broadening is also likely to do strong dephasing due to free electrons and holes in the system. When the system reaches 235 mW, however, a second spectral narrowing is seen. This transition corresponds to standard lasing, and the energy of the photons corresponds to the bare cavity photon energy. At all high powers, the coherent emission stays at this energy, pinned at the cavity photon energy as expected for a laser. 

We thus see that the two different transitions, polariton condensation and standard lasing, are easily distinguishable experimentally, leading to two separate line narrowings at different excitation powers in the same place on the sample. Bloch and coworkers \cite{bloch2thres} have also seen similar behavior by varying the laser power. In general, in all these experiments, it is quite easy to see when the system has reverted to lasing, because the emission energy pops up to the bare cavity photon energy.

\section{Josephson junctions, phase locking, solitons, and vortices}

In the past few years, the number of polariton condensate experiments has exploded, with a number of demonstrations of effects analogous to those of atomic condensates, superconductors or liquid helium. In one sense, it could be said that the main interest of these effects in atomic condensates, superconductors, and helium is to show wave-like behavior of matter; i.e., these effects all follow from being able to write down a Gross-Pitaevskii equation which treats the condensate as a classical wave. In the case of polaritons, the wave-like behavior is perhaps less surprising, because light is already seen as a wave. But unless one disbelieves quantum mechanics, all matter is described as waves. The significance of condensation, with atoms, Cooper pairs, and polaritons, is that a system which normally has strong dephasing can spontaneously acquire a macroscopic coherent amplitude due to the collective effects. Polariton systems, like atomic systems, have a threshold of low temperature and high density at which the coherence due to interparticle interaction defeats the dephasing processes. 

\begin{figure}
\begin{center}
\includegraphics[width=0.75\textwidth]{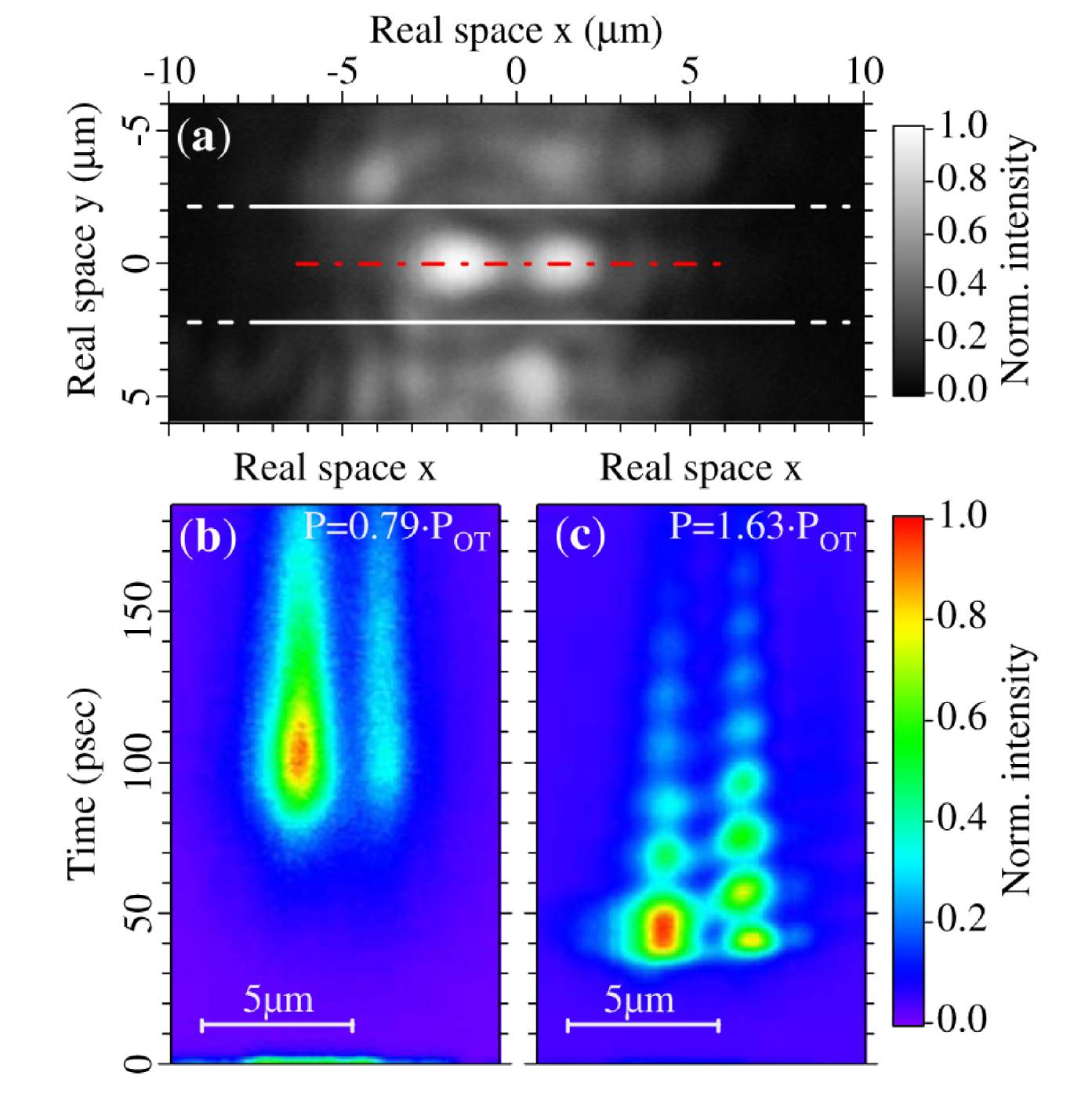}
\end{center}
\caption{Polariton condensate Josephson junction experiment, from Ref.~\protect\cite{devJJ}. a) Spatial intensity distribution for two adjacent traps. b) Time- and space-resolved intensity showing the population of the polaritons in the traps, below the critical density threshold for condensation. c) Time- and space-resolved intensity when the two traps hold polariton condensates. The experiment gives an AC current for a DC potential difference between the two traps.}
\label{fig.JJ}
\end{figure}
Some of these effects with polaritons have analogs in lasing systems, but their behavior is different. Figure~\ref{fig.JJ} shows an example of oscillations in the emission of two traps containing polariton condensates, connected by a thin tunneling barrier. Similar results have been seen by Bloch and coworkers \cite{blochJJ}. 
The period of the oscillations depends on the chemical potential difference between the two traps, as in a standard Josephson junction experiment. 

These oscillations look superficially like the beating of two laser modes. But in the case of laser mode beating, the two states which interfere have discrete energies well separated in frequency from other laser modes. The stability of the oscillations comes from the fact that there are no other nearby states which have gain. In the case of a polariton Josephson junction, as discussed above in Section~\ref{sect.diff}, the stability comes from the interaction of the particles in a continuum of $k$-states of the polaritons.

At first, it may seem that the Josephson junction effect in a superconductor involves very different physics from a Josephson junction of polaritons. The superconductor case involves two Fermi levels of the electrons, and the oscillation frequency depends on the difference between these Fermi levels. But recall that a BCS superconductor is just the high-density limit of BEC (see, e.g., Ref.~\cite{snokebook}, Section 11.2.3). The Fermi level in a BCS superconductor is determined by the pair wave function of the condensed pairs. A condensate of excitons or polaritons also has a spread of $k$-states which is determined by the pair wave function, but the density is low enough that there is no phase space filling which produces a Fermi level in this case. In both cases, once a condensate of pairs is formed, the oscillation frequency of a Josephson junction depends on the chemical potential between the two sides, which depends on the sum of any externally applied potential plus the potential energy due to interactions of the particles. 

Because the chemical potential difference between the two sides depends not just on the static frequency difference between the two traps, but also on the polariton-polariton interactions and on interactions of polaritons with excitons in higher energy states, the period of the oscillations changes in time for the polariton condensate  As the density drops due to recombination, the strength of these interactions drops, and the frequency difference between the traps is shifted. Again, one could argue that a laser system with a strong nonlinear shift of the index of refraction could also show this effect, but as discussed in Section~\ref{sect.GP}, the exceedingly high value of the nonlinear term in the condensate Gross-Pitaevskii equation is what makes the polariton system special. This high value of the nonlinear term comes from dressing the photons in the system with an excitonic part, which gives a strong interaction between the particles through the long-range Coulomb interaction.  The cost of this is that the excitonic part also leads to much higher intrinsic dephasing due to scattering with phonons and impurities. But the bosonic stimulated scattering of the polaritons can overcome this dephasing, leading to phase coherence which is like that of a laser but which shows nonlinear effects at much lower carrier densities. 

There are many other experiments with polaritons that can be described by the formalism of condensation, including vortices \cite{vort1,vort2}, soliton propagation over long distances \cite{soliton}, and propagation without scattering in wires (channels) \cite{blochnew}. The bottom line is that these systems are easily described by the language of condensation, especially when taking into account the interplay between the condensate and excited states in the continuum \cite{haug-g2}, and insisting on the language of laser optics only adds difficulty. The condensation paradigm has become dominant because it is successful in describing the experiments.

When the polariton condensate is pumped resonantly by direct coupling to an external laser beam, the choice of language is more ambiguous. As we have seen, once a condensate is formed, it can be described by a classical wave equation, the Gross-Pitaevskii equation. One can also create such a state by direct pumping with a laser at the same frequency as a polariton state.  In this case the system can be far from equilibrium, and the contribution to the optical effects from particles in incoherent excited states can be unimportant. In general, many of the characteristics of an equilibrium condensate, such as the linear Bogoliubov excitations, persist even as the system moves into a not-fully-equilibrated state \cite{assman}.

\section{``Photon condensation''}
 
\begin{figure}
\begin{center}
\includegraphics[width=0.75\textwidth]{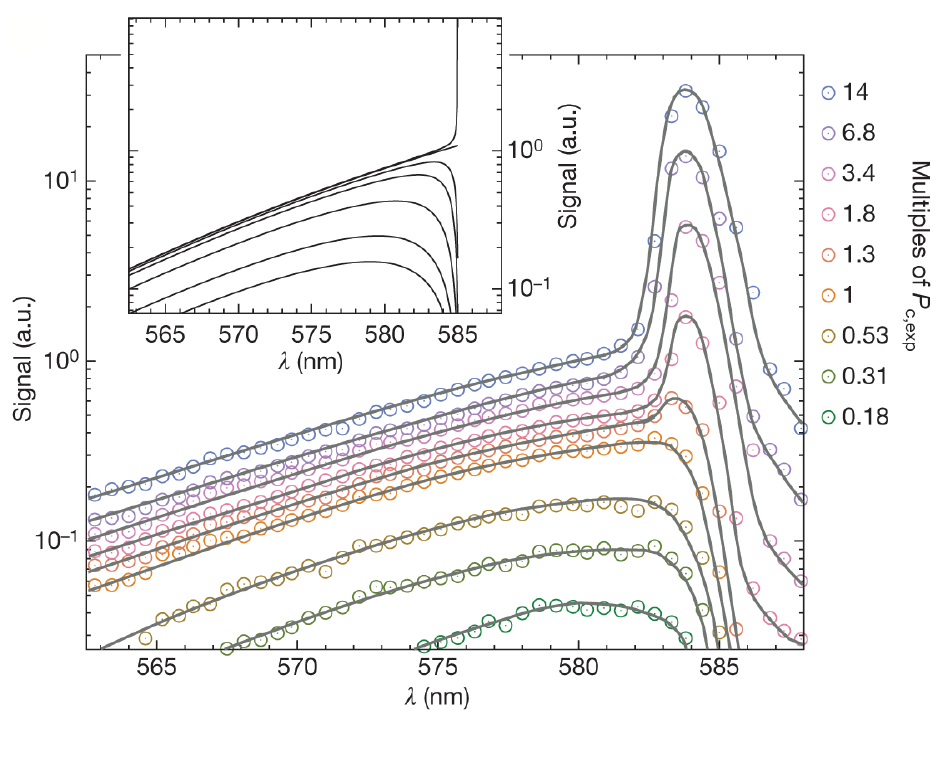}
\end{center}
\caption{Emission energy spectra from a photon gas in weak coupling as a function of density, with a ``photon condensate'' at high density, from Ref.~\protect\cite{photonBEC}. Inset: the prediction for an ideal Bose gas at the same densities.}
\label{fig.pBEC}
\end{figure}

Recently, experiments on ``Bose condensation of photons'' have been reported \cite{photonBEC}. What the authors mean by this is that they created a system with approximate number conservation of photons in a cavity over a short time, similar to the case of a polariton condensate in a microcavity, but in the limit of weak coupling, so that the particles were purely photons. In this limit, the photons thermalized not by collisions with each other, since they are effectively non-interacting, but instead by being absorbed and then re-emitted by dye molecules in the cavity. The dye was chosen such that one photon absorbed led to one photon emitted at a different energy, with very little nonradiative loss; the energy of the emission depended on the temperature of the dye. As in the case of strong coupling, the photons have an effective mass determined by the cavity properties. At high photon density, the energy distribution of the photons thermalized and then showed a peak at $k=0$ as in the case of a polariton condensate, as shown in Fig.~\ref{fig.pBEC}.
 
This system is not the same as standard lasing, because the number of photons was approximately conserved. However, it is not really the same as a standard condensate, either. As discussed above, one of the main properties of standard condensation is spontaneous phase coherence of the condensate, which comes about via interactions of the particles. In the case of the ``photon condensate,'' there is no interaction of the particles which could lead to phase locking. One thus has the situation that the number conservation at a given temperature make the particles want to occupy the lowest available state, but the interaction with the incoherent phonon bath may prevent them from actually becoming a phase coherent condensate.  In Fig.~\ref{fig.pBEC} there is a relatively broad width of the peak at $k=0$, with a full width at half maximum of around 2 nm, or 7 meV, which corresponds to a coherence time of about 100 fs. By contrast, coherence times of hundreds of picoseconds have been observed in polariton condensates \cite{skolT2}. This broad width of the condensate peak in Ref.~\cite{photonBEC} may be limited by the experimental spectral resolution; more recent results \cite{weitzcoh} using interferometry indicate a coherence time on the order of tens of picoseconds.

The experiments with photons in weak coupling are therefore an interesting intermediate case in which the system is prevented from acquiring optical coherence either by lasing or polariton condensation in strong coupling. The system acts in many ways like an ideal Bose gas, which, as is well known, cannot undergo true Bose-Einstein condensation \cite{comb}. 

\section{Conclusions}

Both standard lasing and polariton condensation are states in which there is a coherent polarization of the electrons in the matter, leading to spontaneous phase coherence of the electron polarization and the light emission. In the case of a condensate, the spontaneous coherence comes about due to a thermodynamic phase transition (even when not fully equilibrated) in which temperature is the controlling parameter, while in the case of a laser, the spontaneous coherence comes about far from equilibrium in a transition in which the pump power is the controlling parameter; a laser requires population inversion, while polariton condensation is inhibited by population inversion.  Experiments have been performed in which each of these transitions can be seen separately in the same physical system, with very different behavior.

In many ways the recently reported ``photon condensate'' acts like a polariton condensate, with an effective mass that gives a non-Planckian energy distribution, but the nearly non-interacting nature of the photons in this case means that they can only thermalize via incoherent interactions with an dye medium, which may lead to stronger dephasing. 

{\bf Acknowledgments}. This work has been supported by the National Science Foundation through Grant DMR-0706331.

\end{document}